\documentclass[12pt]{article}
\usepackage{times}
\usepackage{geometry}
\usepackage[utf8]{inputenc}
\usepackage{enumitem,amssymb}
\newlist{thematic}{itemize}{8}
\setlist[thematic]{label=$\square$}
\usepackage{pifont}
\newcommand{\cmark}{\ding{51}}%
\newcommand{\done}{\rlap{$\square$}{\raisebox{2pt}{\large\hspace{1pt}\cmark}}%
\hspace{-2.5pt}}

\usepackage{wrapfig}

\usepackage{natbib}
\usepackage{graphicx}
\usepackage{sectsty}
\sectionfont{\fontsize{13}{13}\selectfont}

\begin{document}

\begin{flushleft}
\huge
Astro2020 Science White Paper \linebreak

\vspace{-0.25cm}

High-Resolution X-ray Imaging Studies of Neutron Stars, Pulsar Wind Nebulae and Supernova Remnants \linebreak

\normalsize
\vspace{-0.35cm}

\noindent \textbf{Thematic Areas:} \hspace*{60pt} $\square$ Planetary Systems \hspace*{10pt} $\square$ Star and Planet Formation \hspace*{20pt}\linebreak
$\done$ Formation and Evolution of Compact Objects \hspace*{31pt} $\square$ Cosmology and Fundamental Physics \linebreak
  $\done$  Stars and Stellar Evolution \hspace*{1pt} $\done$ Resolved Stellar Populations and their Environments \hspace*{40pt} \linebreak
  $\done$    Galaxy Evolution   \hspace*{45pt} $\square$             Multi-Messenger Astronomy and Astrophysics \hspace*{65pt} \linebreak
  
\textbf{Principal Author:}

Name: Samar Safi-Harb
 \linebreak						
Institution: University of Manitoba (Canada)
 \linebreak
Email: samar.safi-harb@umanitoba.ca --
Phone:  (204) 474-7104
 \linebreak
 
 \vspace{-0.3cm}
 
\textbf{Co-authors:}
Elena Amato (INAF-Arcetri, Italy), Eric V. Gotthelf (Columbia, USA), Satoru Katsuda (Saitama, Japan), Manami Sasaki (Erlangen-N\"urnberg, Germany), Yasunobu Uchiyama, Naomi Tsuji (Rikkyo, Japan), Benson Guest (University of Manitoba, Canada)

\end{flushleft}
 
 \vspace{0.1cm}
\noindent 
\textbf{Abstract}:
Supernova remnants serve as nearby laboratories relevant to many areas in Astrophysics, from stellar and galaxy evolution
to extreme astrophysics  and the formation of the heavy elements in the Universe.
The \textit{Chandra} X-ray mission has enabled a giant leap forward in studying both SNRs and their compact stellar remnants on sub-arcsecond scale. However, such high-resolution imaging studies have been mostly limited to the nearby and/or relatively bright objects. There is no question that we are missing a large population, especially in external galaxies. 
Within our own Galaxy, we are presented with new fundamental questions related to neutron stars' diversity, kicks, relativistic winds and the way these objects interact with, and impact, their host environments. In this white paper, we highlight some of the breakthroughs to be achieved with future X-ray missions (such as the proposed \textit{AXIS} probe) equipped with sub-arcsecond imaging resolution and an order of magnitude improvement in sensitivity.

\vspace{0.15cm}

\noindent
\textbf{Endorsed by}:
\footnotesize
Katie Auchettl (University of Copenhagen), 
 Aya Bamba (Univ. of Tokyo),
  Harsha Blumer (West Virginia Univ.),
  Niccol\'o Bucciantini (INAF-Arcetri),
   Yang Chen (Nanjing Univ.), 
  Roger Chevalier (Univ. of Virginia),
  Emma de Ona Wilhelmi
(IEEC-CSIC and DESY),
 Gloria Dubner (Universidad de Buenos Aires),
 Gilles Ferrand (RIKEN),
  Chris Fryer (LANL), 
 Joseph Gelfand (NYU Abu Dhabi),
  Oleg Kargaltsev (George Washington Univ.),
  Noel Klinger (Penn State Univ.),
  Roland Kothes (NRC Herzberg),
 Denis Leahy (Univ. of Calgary),
 Laura Lopez (Ohio State Univ.), 
  Marco Miceli (Palermo),
    Shigehiro Nagataki (RIKEN), 
  Barbara Olmi (Universit\`a di Firenze),
     Salvatore Orlando (INAF-Palermo),
     Sangwook Park (Univ. of Texas at Arlington),
  George Pavlov (Penn State Univ.),
   Paul Plucinsky (Harvard-Smithsonian CfA),
    Adam Rogers (Brandon Univ.),
  Ashley Ruiter \& Ivo Sietenzahl (University of New South Wales),
  Patrick Slane (CfA, Harvard \& Smithsonian),
  Tuguldur Sukhbold (Ohio State Univ.),
 Takaaki Tanaka (Kyoto Univ.), 
 Tea Temim (STSci),
 Diego Torres (IEEC-CSIC, Barcelona),
  Jacco Vink (Univ. of Amsterdam),
 Fr\'ed\'eric Vogt (ESO, Chile),
  Jennifer West (Dunlap Institute),
Brian Williams (NASA GSFC),
George Younes (George Washington Univ.),
Ping Zhou (Univ. of Amsterdam)

\pagebreak

\normalsize

\section{Motivation}
Supernova Remnants (SNRs) are among the most fascinating astrophysical objects in the Universe. They impact the chemical enrichment and evolution of galaxies,  accelerate cosmic rays to extremely high energies, and 
those resulting from core-collapse explosions make the most magnetic and compact objects in the Universe: neutron stars (NSs). NSs  are the best laboratories to study extreme  physical conditions
that can not be achieved even in the most advanced laboratories 
on Earth, as well as relativistic outflows and jets that are ubiquitous in Astrophysics. 
These objects have not only driven scientific breakthroughs, technology development and interdisciplinary connections, but they also fascinate the public and young people.
\\
In this white paper, we focus on SNRs and associated isolated NSs.
We highlight outstanding science breakthroughs that 
can be achieved  \textit{only with sub-arcsecond resolution and high sensitivity}
in the X-ray band -- at least comparable to, or better than, \textit{Chandra}'s imaging resolution,
and with an order of magnitude improvement in sensitivity. 
Such capabilities will be met by the proposed probe \textit{AXIS} \citep{2018SPIE10699E..29M}. \\
We note that the future missions  \textit{Athena} and \textit{Lynx} are expected to achieve 
high-spectral resolution that will significantly benefit SNR science;
however the high-resolution spectroscopy aspect is discussed in separate white papers (B. Williams et al.; L. Lopez et al.). \\
We here specifically aim to address the following questions:
(i) How do pulsars' relativistic winds communicate with, and energize, their surrounding medium?;
(ii) How do NSs evolve and what drives their kicks;
(iii) How do SNRs impact cosmic magnetism and galaxies' evolution?

\section{Neutron Star Winds: How do they impact their surroundings?}

\begin{wrapfigure}{l}{0.6\textwidth} 
\centering
\vspace{-0.0mm}
\includegraphics[scale=0.17]{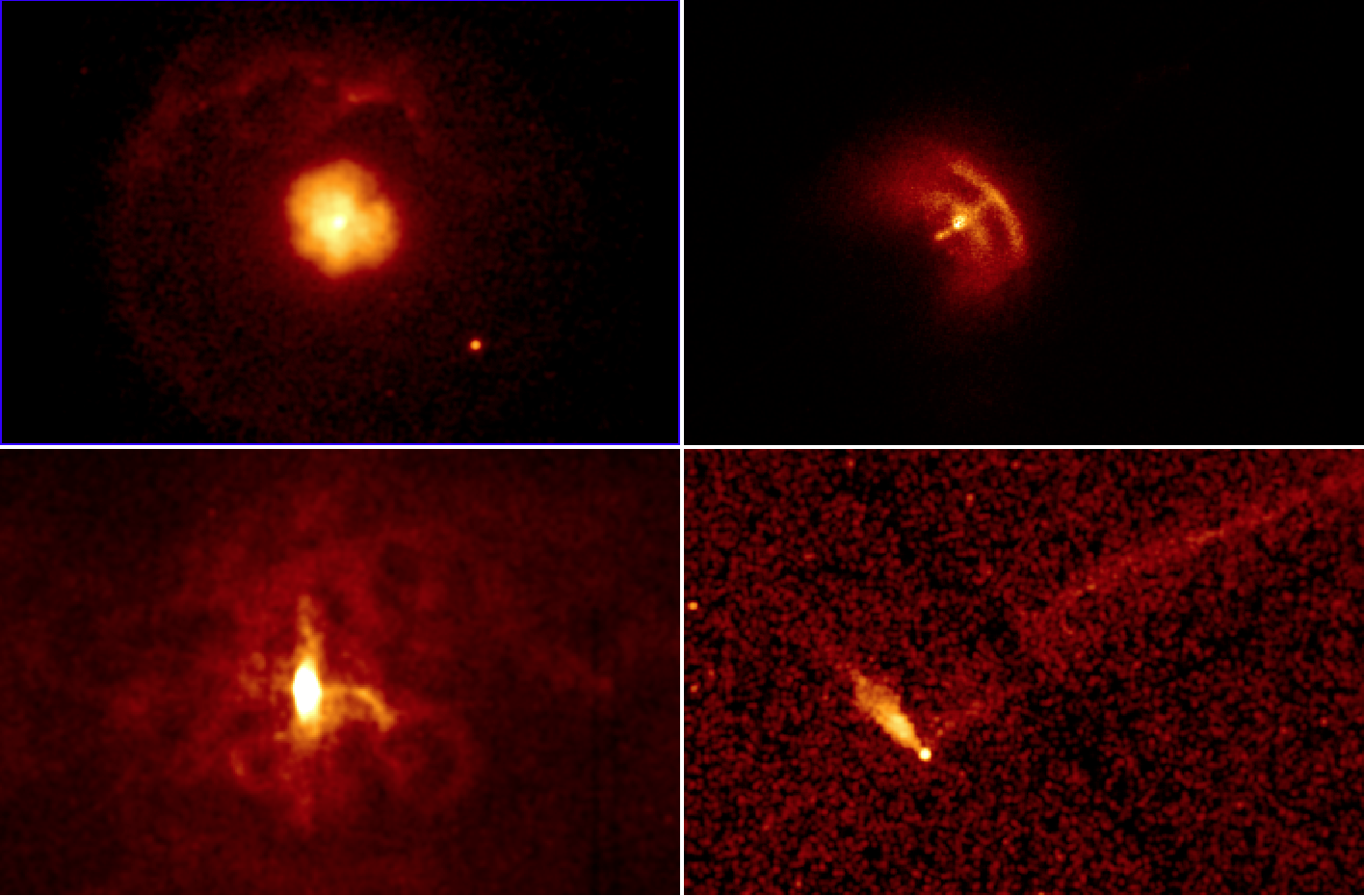}
\vspace{-4.0mm}
\caption{\footnotesize
Simulated \textit{AXIS} images of PWNe, using $Chandra$ images as input, illustrating the fine structures that can be observed with modest exposure times ranging from 25~ks to 100~ks. The high-resolution over a larger FoV and low background will allow the detection of faint, thin structures out to large distances from the powering pulsar.}
\label{fig:PWNe}
\end{wrapfigure}

Magnetized relativistic plasmas phenomena are ubiquious among many classes of astrophysical objects. 
Of these, Pulsar Wind Nebulae (PWNe) \citep{2006ARA&A..44...17G} are some of the best particle accelerators in the Universe, 
with efficiencies close to 30\%, generating particles with energy up to $\sim$1~PeV (cf.  Crab nebula),  the knee in the cosmic ray spectrum.
With its arcsec resolution, \textit{Chandra} opened a new window to resolve torii and jet structures originating from particles accelerated in shocked regions \citep{2008AIPC..983..171K}. Some of these features (resolved also in the optical with HST) are observed to move on sub-arcsecond scales with relativistic velocities. The location of the termination shock, where the ram pressure  of the NS's relativistic wind is balanced by the nebular pressure, often lies $\lesssim$0.1 pc from the pulsar.  
Deep, high resolution AXIS observations offer the possibility of resolving these shocks and providing new insight on the origin, internal dynamics, and evolution of PWNe, and the magnetic field in which they are embedded.

Fig.~\ref{fig:PWNe} shows examples of PWNe simulated with \textit{AXIS} to illustrate the high-resolution structures that can be imaged from young and evolved nebulae using exposure times 5-10 times shorter than \textit{Chandra}'s.
Such high-resolution imaging studies have been mostly limited to the nearby, brightest, or youngest objects; or else very deep exposures are needed for resolving faint structures.
\textit{AXIS} will make a leap in probing a much larger sample of objects in our Galaxy and beyond. 

PWNe in the Magellanic clouds are rare but will be finally within reach. 
An \textit{AXIS}-like resolution of 0.3$^{\prime\prime}$ corresponds to 0.1~pc at the LMC distance,
exactly the scale of termination shocks in PWNe. Increasing the population of spatially resolved PWNe will shed light on star formation as a function of metallicity, but will also answer open fundamental questions about PWNe, pulsars and high-energy astrophysics processes;
some of which are listed here for their broad implications:
\begin{itemize}
\vspace{-0.2cm}
\item{Pulsar (PSR) wind magnetization and anisotropy: These properties of the PSR wind show directly in the appearance of the nebulae, when imaged at high energy with sufficiently high spatial resolution: high speed jets can only appear if the wind is sufficiently magnetized and anisotropic \citep{2004A&A...421.1063D}.
On the other hand, these properties of the wind tell us about the inner workings of the PSR magnetosphere and about magnetic dissipation in relativistic plasmas, a topic which is relevant for many high energy sources and phenomena.}
\vspace{-0.2cm}
\item{
Particle acceleration at a transverse relativistic shock: 
High spatial resolution X-ray imaging and the study of time variability of small-scale features in the inner part of PWNe (X-ray rings and wisps) can constrain where and in what physical conditions particle acceleration occurs 
\citep{2014MNRAS.438.1518O}
and assess whether it is the shock or some other form of dissipation, like magnetic reconnection, that can provide such efficient acceleration.
In addition, the process behind particle acceleration bears information on the pair multiplicity of the PSR magnetosphere, namely the number of electron-positron pairs produced by each electron extracted from the star surface. 
NSs and PWNe are in fact likely to be the primary contributors of the so-called positron excess \citep{2013PhRvL.110n1102A}, that in recent years has attracted much attention both in the cosmic rays and dark matter communities.}
\vspace{-0.2cm}
\item{
Pulsar contribution to leptonic cosmic-rays: 
In bow-shock PWNe (BSPWNe), after a high-speed PSR has left its parent SNR, the wind is only confined by the ISM ram-pressure so that electrons and positrons are free to leave the system and be released in the ISM in the back of the shock, hence contributing to the cosmic ray flux \citep{2018AdSpR..62.2731A}. Details of the particle release are important to determine the PWN contribution to the above-mentioned positron excess. 
Mapping the spectral index close to the NS and out to large distances is essential to assess the particle acceleration process and particle aging effects. Furthermore, multi-TeV particles are sometimes released from the head of the system, through very long and thin X-ray bright channels \citep{2017JPlPh..83e6301K, 2019MNRAS.485.2041B}. \textit{AXIS} will be essential to establish the particle release from BSPWNe and properly synthesize the spectrum of cosmic ray leptons.}
\vspace{-0.2cm}
\item{Magnetar Wind Nebulae: 
There is growing evidence of compact and faint PWNe associated with magnetars, or magnetars-in disguise, whose powering 
Their analysis is complicated by their faintness, compactness and contamination from a dust scattering halo \citep{2013IAUS..291..251S}, \citep{2016ApJ...824..138Y}, \citep{2017ApJ...850L..18B}. 
It is not clear if their X-ray emission is powered by rotation, magnetism or both (e.g.,
\citep{2017ApJ...835...54T}).
\textit{AXIS} will play an important role in studying the nature and origin of nebulae around systems that display a magnetar like activity, thus also requiring a rapid response to bursting sources.}

\end{itemize}

\section{Neutron Stars Diversity: Nature or Nurture?}

The combination of high spatial resolution, low background and
high effective area is ideal for detecting
a new generation of fainter young pulsars (PSRs) in SNRs.
This can help address 
important questions on the birth and evolution of PSRs, and
explain their seemingly diverse properties. 

PSRs were originally detected by their radio emission (e.g., Crab, Vela) but 
X-ray observations over the past two decades have
discovered young, isolated NSs with spectral and timing properties
markedly different from those of the typical rotation-powered radio PSRs.
The radio-quiet PSRs are best described by their implied magnetic
field that range from $10^{10}-10^{14}$~G. The largest of
these are associated with the magnetars, slow rotators ($\sim$2-12 s), that
display a variety of temporal phenomena, such as short and long
scale transient outbursts, random episodes of short ($\sim$1~s) burst of
hard X-rays, and erratic spin-down.  Most notably, their X-ray
emission far exceeds that predicted for a rotation-powered PSR,
based on their spin-down luminosity.
X-ray emission of magnetars is believed to arise from magnetic losses from a strongly magnetized
($B>4.4\times10^{13}$~G) isolated NS \citep{1992ApJ...392L...9D}.
However, the detection of magnetar-like activity from seemingly
classical rotation-powered PSRs \citep{2008Sci...319.1802G,2008ApJ...678L..43K, 2016ApJ...829L..25G}
complicates this picture. 
Similarly, pulsations detected from the central compact objects (CCOs), PSRs in SNRs with extremely
small magnetic dipole fields \citep{2013ApJ...765...58G}, are a puzzle 
since the only mechanism thought to be capable of creating a non-uniform surface temperature is anisotropic
heat conduction in a strong  magnetic field.

The latter problem can be addressed with \textit{AXIS} spectral-temporal
observations of young NSs to model their surface emission using
phase-resolved spectroscopy. This will allow a better understanding of
their magnetic field configurations. A leading explanation for the
generation of the observed hot spots on CCOs requires crustal toroidal fields
to insulate the magnetic equator from heat conduction. This toroidal
component is expected to be generated by differential rotation in the
proto-neutron star dynamo.  To have a significant effect on the heat
transport, the crustal toroidal field required in all models
is $>10^{14}$~G, far greater than the poloidal
field, if the latter is measured by the spin-down.  Do all CCOs harbor
an inner magnetar buried in their crust without (currently)
contributing to its external dipole responsible for its slow spin-down?
Most importantly, how do the CCOs and the magnetars relate to each
other and to the classical rotation-powered radio PSRs?

The proposed very high spatial resolution of \textit{AXIS} coupled with its high time
resolution CCD imaging modes can provide breakthrough science for
faint CCOs in SNRs. Of great interest is detecting pulsations from the
300~year-old CCO in SNR Cassiopeia~A, the long-sought compact object discovered
with the first light \textit{Chandra} observation \citep{2000ApJ...531L..53P}. 
This will provide the critical energetics and magnetic field
estimates for a PSR close to its birth values.  Alternatively, a
strong upper limit on any pulsations would help advance theories of
atmospheric physics of NSs. Lastly, the order of magnitude increase
in sensitivity will be needed to discover the missing CCO decendants, 
and in turn address the population of core-collapse SNRs in our Galaxy.

 \section{Neutron Star Velocities: What drives their kicks?}

The origin of high velocities in NSs is a long-standing mystery in astrophysics.  
There are two main competing mechanisms to kick NSs: (a) anisotropic ejection of the stellar debris (`hydrodynamic kick', \cite{1994A&A...290..496J})
and (b) asymmetric-neutrino emission (`neutrino-induced kick', \cite{1993A&AT....3..287B}).
Fortunately, the two scenarios predict a clear difference in NS kick velocities and SN asymmetries.  The hydrodynamic kick mechanism predicts that NS velocities are directed opposite to the stronger explosion where explosive nucleosynthesis elements from Si to Fe are preferentially expelled, whereas the neutrino-induced kick mechanism either suggest no correlation between NS velocities and SN asymmetries, or predict the strongest mass ejection in the direction of NS motion.  
Recent X-ray observations of SNRs revealed that NSs preferentially move opposite to the bulk of either X-ray emission \citep{2017ApJ...844...84H} or intermediate-mass elements \citep{2018ApJ...856...18K},
supporting the hydrodynamic kick scenario.  However, in many cases, NS velocities are indirectly inferred from displacements between NS positions and geometric centers. 
The number of robust samples is still quite small.  An instrument like \textit{AXIS}
will significantly increase the observational sample, as described below.

\noindent To distinguish between the two NS kick scenarios, it is critically important to measure both NS proper motions and detailed distributions of SN ejecta.  NS proper motions are important not only because they allow us to estimate NS velocities, but also because they help to check if a NS is really associated with a SNR and to infer an explosion site by tracing back the proper motion.  The long time baseline, which will be available with the combination of \textit{Chandra} and \textit{AXIS}, will be the key to reduce systematic uncertainties on NS proper motions.  In addition, AXIS's superior throughput and wide field of view, together with its moderate spectral resolution, will allow us to map detailed SN ejecta distributions for faint and large SNRs.

So far, we have only three SNRs (Puppis A, Cas~A, G292.0+1.8) for which both NS kick velocities and ejecta distributions are robustly estimated.  
All three systems show an anti-correlation between NS kick velocities and ejecta distributions, favoring the hydrodynamic origin for the NS kicks.  \textit{AXIS} will increase the number of such samples substantially, 
and will reveal if the hydrodynamic kick scenario is the only process that can accelerate the NS or 
if other mechanisms, such as the neutrino-induced kick scenario, can play a role.
Increasing the samples is also important to search for correlations between the degrees of explosion asymmetries and NS kick velocities, and between the NS surface magnetic fields and NS kick velocities.

\section{SNR shock impact on cosmic magnetism}

\begin{wrapfigure}{l}{0.5\textwidth} 
\vspace{-4.5mm}
\includegraphics[scale=0.4]{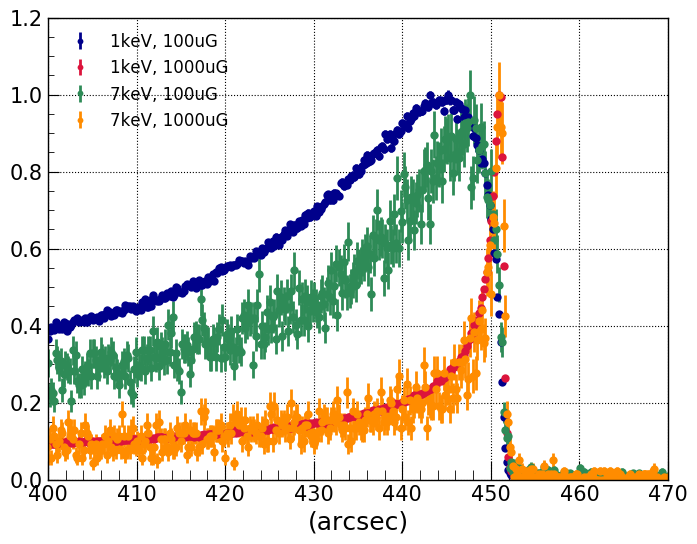}
\vspace{-4.5mm}
\caption{\footnotesize 
Simulated radial profiles of shocked emission from a young, 20 pc diameter SNR, at a distance of 1 kpc.
The 100 ks AXIS simulations, for an amplified magnetic fields of 100$\mu$G and 1~mG, illustrate that 
the shock  narrows in width for stronger fields.}
\label{fig:Bshock}
\end{wrapfigure}

\vspace{-0.3cm}
Turbulent magnetic fields at young SNR shocks are expected to be significantly amplified by a cosmic-ray current driven instability that develops in the shock precursor. Magnetic field amplification (MFA) is thought as the key element in non-linear Diffusive Shock Acceleration theory \cite{2014ApJ...789..137B}. 
X-ray observations with {\it Chandra} have revealed the presence of narrow synchrotron X-ray filaments 
at the outer edge of young SNRs, demonstrating that the strong shocks at young SNRs are indeed capable of amplifying the interstellar magnetic field by large factors \cite{2003ApJ...584..758V}. 
The narrowness of synchrotron X-ray filaments could be due to rapid synchrotron cooling of high-energy electrons in the postshock flow if the magnetic field reaches $\sim0.1$~mG. 
In some cases, time-variability of the synchrotron X-rays can be seen, which is another 
evidence in favor of MFA \citep{2007Natur.449..576U}. \\
The turbulent magnetic fields, likely amplified by CR current driven instabilities, can be imprinted in the spatial structures of the synchrotron X-ray filaments. 
Testing the theoretical predictions is best achieved with high-angular resolution measurements of the energy dependence on the width of the filamentary structures (see e.g., \citep{2007A&A...465..695Z}). 
This is illustrated in Fig.~2 showing the radial profiles of the synchrotron X-ray filaments expected to be observed by \textit{AXIS}.

\section{Population Studies: From our Galaxy to the Nearby Universe}
SNRs  radiate copiously at energies of 0.5 -- 2 keV, a range that is difficult 
to study globally in the Milky Way because of absorption by matter in the
Galactic disk, with an absorbing column density that amounts to a few 
10$^{21} \, {\rm cm}^{-2}$ within less than a kpc.
In order to study the population of SNRs in a galaxy as a whole we have
to look beyond the Milky Way.

By studying SNRs in nearby galaxies, we particularly want to address following
questions:
\begin{itemize}
\vspace{-0.2cm}
\item What is the fraction and spatial distribution of core-collapse SNRs
vs. type Ia SNRs? These can be identified based on their morphology combined
with spectral properties.
\vspace{-0.2cm}
\item What is the X-ray luminosity function (XLF) of SNRs? How are the XLFs of different galaxies related to the underlying stellar population, ISM, 
metallicity and SNR evolution?
\vspace{-0.2cm}
\item What is the distribution of SNRs in comparison to that of the cold ISM?
Are SNRs correlated with large structures in the ISM or with star-forming 
regions?
\vspace{-0.2cm}
\item How many of the SNRs show correlations with molecular clouds? Can the
SNR population explain the cosmic ray density in galaxies?
\end{itemize}
First X-ray surveys of the larger galaxies in the Local Group, the Magellanic
Clouds, M31 and M33 were performed with \textit{Einstein} and \textit{ROSAT}, yielding 
catalogs of SNRs and candidates in these galaxies.
A detailed list of X-ray SNRs in M31 was created using an XMM-Newton survey.
While SNRs in the Magellanic Clouds can be generally well resolved spatially and
studied in detail \citep{2016A&A...585A.162M},
so far, only a few SNRs in M\,31 have been resolved with \textit{Chandra}
\citep{2002ApJ...580L.125K,2003ApJ...590L..21K,2004ApJ...615..720W}.

The Magellanic Clouds, M31 and M33 have very different ISM densities,
metallicities, and star formation rates, making differences in
their ensemble of X-ray SNRs of great interest for testing theories for the
dominant SNR and ISM characteristics that cause the X-ray emission of SNRs.
\textit{AXIS} will allow us to extend the study of SNR populations to nearby galaxies 
outside the Local Group including large spiral galaxies like
M81, M83, or NGC~300, in which candidates of X-ray SNRs
have been detected using \textit{XMM-Newton} or \textit{Chandra}, but more detailed studies have
not been possible.
In addition, 0.3$^{\prime\prime}$ corresponds to $\sim$1.1 pc at the distance of M31
which will allow us to detect and resolve all mature SNRs
in M31 and M33, allowing for the first time
detailed X-ray population studies of SNRs in galaxies beyond the Magellanic Clouds.
\\
Last but not least, the combined sub-arcsecond resolution and high sensitivity will be needed to resolve small/young remnants, including SN~1987A which will should enter the ejecta-dominated phase in the 2030's \citep{2016ApJ...829...40F}. This is also crucial for comparing the X-ray emission with that at other wavelengths, where high-resolution images are or will be available in the future.

\vspace{0.1cm}
{\bf In summary:} Following up on the legacy of \textit{Chandra}, an X-ray telescope with sub-arcsecond resolution combined with 
high sensitivity and ToO capabilities will further revolutionize the field of SNRs, PWNe and NSs in our Galaxy and the nearby Universe. These capabilities will be an absolute requirement to advance the field, especially in synergy with upcoming high-resolution facilities across the electromagnetic spectrum. 

\newpage
\bibliographystyle{plain}
\bibliography{references}
\end{document}